\documentclass[aps,prd,11pt]{revtex4}
\usepackage[colorlinks=true, pdfstartview=FitV, linkcolor=blue, citecolor=red, urlcolor=magenta]{hyperref}
\usepackage{graphicx}
\usepackage{latexsym}
\usepackage{amsmath}
\usepackage{amsfonts}
\usepackage{amssymb}
\usepackage{verbatim}

\newcommand{\be}{\begin{equation}}
\newcommand{\ee}{\end{equation}}
\newcommand{\bea}{\begin{eqnarray}}
\newcommand{\eea}{\end{eqnarray}}

\newcommand{\ben}{\begin{eqnarray}}
\newcommand{\een}{\end{eqnarray}}

\begin{document}

\title{Cosmological scenarios in modified gravity with non-dynamical fields}
\author{D. Bazeia$^{a}$, F.A. Brito,$^{b}$ and F.G. Costa,$^{a}$ } 
\affiliation{
$^{a}$Departamento de F\'\i sica,
Universidade Federal da Para\'\i ba, Caixa Postal 5008,
58051-970 Jo\~ ao Pessoa, Para\'\i ba,
Brazil\\
$^{b}$Departamento de F\'\i sica,
Universidade Federal de Campina Grande, Caixa Postal 10071,
58109-970  Campina Grande, Para\'\i ba,
Brazil
}

\begin{abstract}
In this paper we address the issue of exploring some cosmological scenarios in modified Einstein gravity through non-dynamical (auxiliary) fields. 
We found that all scenarios are controlled by a specific parameter associated with an auxiliary field. We explore the emergence of inflationary,  radiation, matter and dark energy dominated regimes. Furthermore, an interesting possibility such as the emergence of a self-tuning mechanism to the cosmological constant problem in the radiation dominated era is also discussed. 
\end{abstract}
\pacs{xxx,yyy} \maketitle


\section{Introduction}

One of the main problems of constructing alternative theories of gravity by adding extra dynamical fields non-minimally coupled to gravity is to evade the presence of extra propagating degrees of freedom which lead to instabilities in these theories \cite{Will:2005va,Woodard:2006nt}. However, as has been shown very recently by Pani, Sotiriou, and Vernieri \cite{Pani:2013qfa}, one may circumvent this problem by   modifications of Einstein gravity with non-dynamical (auxiliary) fields. See also Palatini $f(R)$ and other modified gravities \cite {HAB,Sotiriou:2008rp,DeFelice:2010aj,Nojiri:2010wj,Capozziello:2013xn,Capozziello:2010uv,Bazeia:2014xxa} and Eddington-inspired Born-Infeld theory \cite{banados} for related issues. One of the main consequences of gravity theories with auxiliary fields is that they lead to the presence of higher-order derivatives of the matter fields. In the next-to-leading order in the derivative of matter fields the parametrization of the auxiliary fields is simply restricted to two parameters apart from the cosmological constant. Because of the higher-order derivatives of the matter fields in the field equations, these parameters can be severely constrained due to the response of the metric to the abrupt changes in the matter energy density  \cite{Pani:2013qfa}. In other words this means the presence of undesirable singularities in the theory.  However, in a recent study in Ref.~\cite{Kim:2013nna} by considering Eddington-inspired Born-Infeld theory (which in some approximation can be seen as a special case of the new theory \cite{Pani:2013qfa}) it was pointed out that such singularity can be removed by some mechanism. In this spirit of modified theories of gravity one has already shown in the literature that  Eddington-inspired Born-Infeld is identical to bigravity theory \cite{Delsate:2012ky}. More recently this modified gravity with non-dynamical fields was extended to the thick braneworld model in five dimensions  \cite{Guo:2014bxa} to address the issue of gravity localization. A similar route in bigravity theory was also taken in Ref.~\cite{Bazeia:2012br}. In the following we shall focus our attention to cosmological scenarios.

In this paper we look for cosmological scenarios in this new theory \cite{Pani:2013qfa}.  By considering the modified Einstein equations with a non-dynamical field in the Friedmann-Robertson-Walker background we find the modified Friedmann equations. We concentrate our analysis up to linear modifications ---  very recently appeared a similar study considering higher order auxiliary fields \cite{Harko:2014zma}. We show that even at this level the results are far from being trivial. The modified equation of state gives a richer cosmological scenario with several dominated regimes. The existence of the new non-dynamical field allows for dark energy in the modified theory even if the equation of state of the unmodified theory is just the matter dominated regime. Another interesting point is that in the radiation dominated regime emerges a self-tuning mechanism \cite{1,2,3} to the cosmological constant problem \cite{weinberg}. See also Ref.~\cite{Bazeia:2012vy} for a recent discussion on such mechanism.

The paper is organized as follows. In Sec.~\ref{formalism} we briefly present the formalism of the modified gravity with auxiliary fields. In Sec.~\ref{scenarios} we discuss the possible cosmological scenarios in this new theory of gravity. In Sec.~\ref{scenarios} we present the emergence of a self-tuning mechanism to the cosmological constant  problem in the radiation dominated era. Finally in Sec.~\ref{conclu} we present our final considerations.

\section{The formalism}
\label{formalism}
The field equations for  the modified Einstein equations with auxiliary fields read 
\bea\label{Mod_Einstein}
G_{ab}+\Lambda g_{ab}&=&T_{ab}+S_{ab}
\eea
where \cite{Pani:2013qfa}
\bea\label{correct_T}
S_{ab}&=&\alpha_1 g_{ab}T+\alpha_2g_{ab}T^2+\alpha_3TT_{ab}+\alpha_4T_{cd}T^{cd}+\alpha_5T^c_aT_{cb}+\beta_1\nabla_a\nabla_bT\nonumber\\
&+&\beta_2g_{ab}\square T+\beta_3\square T_{ab}+2\beta_4\nabla^c\nabla_{(a}T_{b)c}+...
\eea

Now we shall keep only non-derivate linear terms in $T$, consider $\Lambda\to0$ and assume $S_{ab}\ll T_{ab}$ in order to maintain the modified Einstein equations (\ref{Mod_Einstein})  divergent free with a non-dynamical field parametrized by $\alpha_1$ as follows 
\bea\label{eom}
G_{\mu\nu}=8\pi G\left[T_{\mu\nu}+\alpha_1Tg_{\mu\nu}\right],\qquad \mu,\nu=0,1,2,3
\eea
with the trace of the energy-momentum given by the usual form $T=\rho-3p$. We also recover the factor $8\pi G$, which is normalized to unit in the original Ref.~\cite{Pani:2013qfa}. Let us now assume the FRW metric (assuming a flat Universe, i.e., $k=0$) 
\bea\label{metric}
ds^2=g_{\mu\nu}dx^\mu dx^\nu = dt^2-a^2(t)d\vec{x}^2
\eea
By using the metric (\ref{metric}) into the Einstein equations (\ref{eom}) we find
\ben\label{friedmann1}
\frac{\dot{a}^2}{a^2}=\frac{8\pi G}{3}\left[(1+\alpha_1)\rho-3\alpha_1p \right]
\eea
and
\ben\label{friedmann2}
\frac{\ddot{a}}{a}=-\frac{4\pi G}{3}\left[(1-2\alpha_1)\rho+(6\alpha_1+3)p \right]
\eea

\section{Cosmological scenarios}
\label{scenarios}

Let us start with Eqs.~(\ref{friedmann1})-(\ref{friedmann2}) and the equation of state $p=\omega\rho$ to rewrite them as following 
\ben\label{friedmann1.0}
H^2=\frac{8\pi G}{3}\left[(1+\alpha_1)-3\alpha_1\omega \right]\rho
\eea
and
\ben\label{friedmann2.0}
\dot{H}+H^2=-\frac{4\pi G}{3}\left[(1-2\alpha_1)+(6\alpha_1+3)\omega \right]\rho
\eea
where $H=\dot{a}/a$ is the Hubble parameter. Differentiating Eq.~(\ref{friedmann1.0}) with respect to $t$ we find the relationship between time derivative of energy density and Hubble parameter 
\bea\label{hubble1}
3H\dot{H}=4\pi G\left[(1+\alpha_1)-3\alpha_1\omega\right]\dot{\rho}
\eea
Another important relationship between these quantities can be found by substituting Eq.~(\ref{friedmann1.0}) into Eq.~(\ref{friedmann2.0}) which reads
\bea\label{hubble2}
\dot{H}=-4\pi G(1+\omega)\rho
\eea
Now combining the Eqs.~(\ref{hubble1})-(\ref{hubble2}) we find the following important differential equation 
\bea\label{energy}
\dot{\rho}+3H\eta\rho=0,\qquad \eta=\left[\frac{(1+\omega)}{(1+\alpha_1)-3\alpha_1\omega} \right]
\eea
where $\eta\to(1+\omega)$ in Einstein gravity ($\alpha_1\to0$). For the sake of comparison, the modified equation of state in our scenario is defined as $\omega_\eta=\eta-1$. By solving the differential equation (\ref{energy}) we find the standard solution
\bea\label{rho}
\rho(t)=\frac{\rho_0}{a^{3\eta}(t)}
\eea
In order to find explicit solutions for $a(t)$ one should substitute Eq.~(\ref{rho}) into Eq.~(\ref{friedmann1.0}). Finally, we find the form of $a(t)$ as a function of $t$ given by
\bea\label{sol_a}
a(t)=a_0\,t^{\frac{2}{3{\eta}}},\qquad a_0=\left(\frac{3\eta}{2}\right)^{\frac{2}{3\eta}}\left[\frac{8\pi G}{3}\Big((1+\alpha_1)-3\alpha_1\omega\Big)\rho_0\right]^{\frac{1}{3\eta}}
\eea
The density as a function of $t$ can be readily found from Eqs.~(\ref{rho})-(\ref{sol_a}). The explicit form reads
\bea
\rho(t)=\frac{\rho_0}{a_0^{3\eta}}\frac{1}{t^2}
\eea
As usual, this implies that the Hubble parameter $H^2\sim \rho$ scales as $H\sim 1/t$. As well-known, this, however, is not true for all possible cosmological scenarios. There is an important exception. In the vacuum dominance the equation of 
state is $\omega=-1$. In this sense from equation (\ref{energy}) we see that $\eta=0$ and $\dot{\rho}=0$, which implies that $H^2\sim\rho=\rho_0\equiv const.$ The expansion in this case is exponential and the solution
(\ref{sol_a})  is replaced by something like $a(t)\sim\exp{[\rho_0^{1/2}t]}$.

More precisely, we should address the vacuum scenario separately. By substituting the equation of state $p=-\rho$, recalling that $\rho=\rho_0$, into Eqs.~(\ref{friedmann1})-(\ref{friedmann2}) we find
\ben\label{friedmann1.3}
\frac{\dot{a}^2}{a^2}=\frac{8\pi G}{3}\left(1+4\alpha_1 \right)\rho_0
\eea
and
\ben\label{friedmann2.3}
\frac{\ddot{a}}{a}=\frac{8\pi G}{3}\left(1+4\alpha_1 \right)\rho_0
\eea
or simply 
\bea\label{aaa}
\frac{\dot{a}^2}{a^2}=\frac{\ddot{a}}{a}
\eea
By using any of these equations we should find the aforementioned exponential solution
\bea
a(t)=a_0\exp{\left(\sqrt{\frac{8\pi G}{3}(1+4\alpha_1)\rho_0}\,t \right)}
\eea
The above analysis simply shows that the accelerating regime governed by the exponential expansion (due to $\omega=-1$) takes places when $\alpha_1>-1/4$. The case $\alpha_1<-1/4$ is consistent with oscillatory solutions for non-flat Universe, i.e., $k\neq0$ into (\ref{friedmann1.3}) --- see Ref.~\cite{Cai:2010zma,Cai:2011bs} for cyclic cosmology in a similar context. The regime for $\alpha_1=-1/4$ is special and will be discussed with further detail in the next section. In the following we shall address the issues of some special cosmological scenarios.

The {\sl matter dominated} regime develops under the power law $\eta=1$ into (\ref{rho}). This gives the relation $\omega=\alpha_1/(3\alpha_1+1)$. For $\alpha_1=0$ we find the usual solution $\omega=0$, as we can easily see from Eqs.~(\ref{energy}), (\ref{rho}) and (\ref{sol_a}). If the parameter $\alpha_1$ runs for large enough values we find $\omega\to 1/3$, which mimics the {`radiation dominated'} equation of state but with matter dominated behavior. On the other hand, for $3\alpha_1\ll1$ we find $\omega=\alpha_1\ll1/3$ which may be related to {\sl dark matter dominated} regime.

However, the {\sl radiation dominated} regime develops under the power law $\eta=4/3$ into (\ref{rho}). Interestingly enough, in this case gives a unique solution $\omega=1/3$. In this case the modified equation of state $\omega_\eta=\eta-1=4/3-1=\omega=1/3$ coincides with the equation of state of radiation dominated regime of the Einstein gravity. We have more to say about this point in the next section.

The {\sl dark energy dominated} regime shows up for $\omega=0$ and $\alpha_1\neq0$, for instance. We can see from Eq.~(\ref{sol_a}) that an accelerated regime is possible for $2/3\eta>1$. This is accomplished as long as $\alpha_1>1/2$. Suppose $\alpha_1=1$ then the solution is
\bea
a(t)=a_0\,t^{4/3}\sim t^{1.3}
\eea
In summary, the above and several other regimes such as {\sl stiff fluid} and {\sl phantom cosmology} can also be easily explored by using the important parameter $\eta$ given in Eq.~(\ref{energy}) or more explicitly by using the modified equation of state
\bea
\omega_\eta=\eta-1=\frac{\omega-\alpha_1+3\alpha_1\omega}{1+\alpha_1-3\alpha_1\omega}
\eea

\section{A self-tuning mechanism to cosmological constant problem}
\label{tuning}

It is worth noting that from Eqs.~(\ref{friedmann1})-(\ref{friedmann2}) one can obtain a self-tuning mechanism to address the cosmological constant problem, at least in a specific phase of the cosmological evolution. According to Eqs.~(\ref{friedmann1.3})-(\ref{aaa}), setting $\alpha_1=-1/4$ we exclude vacuum dominance in the cosmological scenarios. Furthermore, with this choice
we find the modified Friedman equations \cite{3}
\ben\label{friedmann1.1}
\frac{\dot{a}^2}{a^2}=\frac{8\pi G}{3}\left(\frac34\right)\left(\rho+p \right)=2\pi G \left(\rho+p \right)
\eea
and
\ben\label{friedmann2.1}
\frac{\ddot{a}}{a}=-\frac{4\pi G}{3}\left(\frac32\right)\left(\rho+p \right)=-2\pi G \left(\rho+p \right)
\eea
Again, even if the vacuum contribution comes from the matter sector, that is, for example, if the species are distributed according to $p=p_\Lambda+p_{radiation}+p_{matter}$ and $\rho=\rho_\Lambda+\rho_{radiation}+\rho_{matter}$, being $p_\Lambda=-\rho_\Lambda$ the equation of state of the vacuum, then no one of the above equations can `see' this vacuum contribution.
Now comparing Eqs.~(\ref{friedmann1.1})-(\ref{friedmann2.1}) we find
\bea 
\frac{\dot{a}^2}{a^2}=-\frac{\ddot{a}}{a}
\eea
This equation is satisfied by the solution describing the phase of the radiation of the Universe i.e.,
\bea\label{sol}
a(t)\sim t^{1/2}
\eea
This is not surprising since both equations (\ref{friedmann1.1})-(\ref{friedmann2.1})  are consistent with the equation of state $p=\omega \rho$  for radiation $\omega=1/3$ in the Einstein gravity.
Let us explore the solution (\ref{sol}) as follows. By substituting (\ref{sol}) into Eq.~(\ref{friedmann1.1}) we find
\ben\label{friedmann1.2}
2\pi G \left(\rho+p \right)=\frac{\dot{a}^2}{a^2}\sim\frac{1}{t^2}
\eea
that using the equation of state for radiation, i.e., $p=(1/3)\rho$ one finds 
\ben\label{friedmann1.4}
\rho=\frac{\rho_0}{a^4(t)}
\eea
Interestingly enough, notice that  for $\alpha_1=-1/4$ into Eq.~(\ref{energy})  there is no contribution of the original equation of state $\omega$. In this particular case, we always have $\eta=4/3$ and Eq.~(\ref{rho}) coincides with Eq.~(\ref{friedmann1.4}). We conclude that fixing the auxiliary parameter as $\alpha_1=-1/4$ we naturally have a cosmological scenario in the radiation regime. Furthermore, no vacuum contribution  is present in this regime and no cosmological constant issues appear during this phase.

\section{Conclusions}
\label{conclu}

In summary we have considered the recently introduced modified gravity theory through auxiliary fields. This theory does not present undesirable extra degrees of freedom, although some singularities may appear due to the higher order derivative matter fields. However this seems not to be a problem since some mechanism to solve this problem has been proposed in Eddington-inspired Born-Infeld theory which can be seen as a special case of this new theory under some approximation. In the present study, in order to address the cosmological scenarios, we considered just the linear modifications, which already  revealed to be able to develop a richer cosmological scenario. We have identified an interesting emergence of a self-tuning mechanism to the cosmological constant issue in the radiation dominated regime. For future investigations it should be interesting to consider higher order modifications to see how such a mechanism works. 

{\acknowledgments} We would like to thank to CNPq, PNPD-CAPES, PROCAD-NF/2009-CAPES for partial financial support.

\end{document}